\documentclass[12pt]{article}
\usepackage{axodraw,bbold}

\catcode`@=12
\begin{document}
\thispagestyle{empty}
\begin{flushright} 
UCRHEP-T436\\ 
September 2007\
\end{flushright}
\vspace{0.5in}
\begin{center}
{\LARGE	\bf Z$_3$ Dark Matter and\\ Two-Loop Neutrino Mass\\}
\vspace{1.5in}
{\bf Ernest Ma\\}
\vspace{0.2in}
{\sl Department of Physics and Astronomy, University of 
California,\\ Riverside, California 92521, USA\\}
\vspace{1.5in}
\end{center}

\begin{abstract}\
Dark matter is usually distinguished from ordinary matter by an odd-even 
parity, i.e. the discrete symmetry $Z_2$.  The new idea of $Z_3$ dark 
matter is proposed with a special application to generating radiative 
Majorana neutrino masses in two-loop order.
\end{abstract}

\newpage
\baselineskip 24pt

Neutrinos have mass \cite{gm07} and the Universe has dark matter \cite{bhs05}. 
Whereas the nearly massless neutrinos themselves could only account for a 
very small fraction of the latter, the mechanism by which they acquire mass 
may involve particles which do form the bulk of the dark matter itself. 
A recent realistic proposal \cite{m06-1,kms06,m06-2,m06-3,hkmr07,ms07,bm07} 
is to add a second scalar doublet $\eta = (\eta^+,\eta^0)$ to the Standard 
Model (SM) of particle interactions together with three singlet neutral 
Majorana fermions $N_i$, such that $\eta$ and $N_i$ are odd under an exactly 
conserved $Z_2$ discrete symmetry whereas all SM particles are even.  In the 
presence of the allowed quartic scalar interaction
\begin{equation}
{1 \over 2} \lambda_5 (\Phi^\dagger \eta)^2 + H.c.,
\end{equation}
where $\Phi = (\phi^+,\phi^0)$ is the SM Higgs doublet, $\eta^0$ is 
split so that $\eta^0_R = \sqrt{2} Re(\eta^0)$ and $\eta^0_I = \sqrt{2} Im 
(\eta^0)$ have different masses, resulting simultaneously in (A) the one-loop 
radiative generation (Fig.~1) of Majorana neutrino masses through the allowed 
interaction
\begin{equation}
h_{ij} (\nu_i \eta^0 - l_i \eta^+) N_j + H.c.,
\end{equation}
and (B) the possible identification \cite{m06-1,bhr06,lnot07,glbe07} of 
$\eta^0_R$ or $\eta^0_I$ as dark matter.  The collider phenomenology of 
$\eta$ has also been discussed \cite{bhr06,cmr07}.

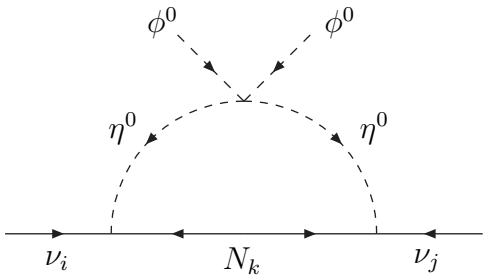
\begin{figure}[htb]
\begin{center}
\begin{picture}(360,120)(0,0)
\ArrowLine(90,10)(130,10)
\ArrowLine(180,10)(130,10)
\ArrowLine(180,10)(230,10)
\ArrowLine(270,10)(230,10)
\DashArrowLine(155,85)(180,60)3
\DashArrowLine(205,85)(180,60)3
\DashArrowArc(180,10)(50,90,180)3
\DashArrowArcn(180,10)(50,90,0)3

\Text(110,0)[]{$\nu_i$}
\Text(250,0)[]{$\nu_j$}
\Text(180,0)[]{$N_k$}
\Text(135,50)[]{$\eta^0$}
\Text(230,50)[]{$\eta^0$}
\Text(150,90)[]{$\phi^{0}$}
\Text(217,90)[]{$\phi^{0}$}

\end{picture}
\end{center}
\caption{One-loop generation of neutrino mass.}
\end{figure}

The simple idea that $\eta^0_R$ or $\eta^0_I$ is absolutely stable because of 
the $Z_2$ discrete symmetry goes back 30 years \cite{dm78}.  On the other 
hand, the simplest possible realization of dark matter is to postulate 
\cite{sz85} a real scalar field $D$ which is odd under $Z_2$.  The SM 
Higgs potential is then extended to
\begin{equation}
V = m_1^2 \Phi^\dagger \Phi + {1 \over 2} m_2^2 D^2 + {1 \over 2} \lambda_1 
(\Phi^\dagger \Phi)^2 + {1 \over 8} \lambda_2 D^4 + {1 \over 2} \lambda_3 
D^2 (\Phi^\dagger \Phi).
\end{equation}
In the presence of electroweak symmetry breaking, $\Phi^\dagger \Phi$ is 
replaced by $(v+h)^2/2$, where $m_1^2 + \lambda_1 v^2/2 = 0$, hence
\begin{equation}
V = {1 \over 2} \lambda_1 v^2 h^2 + {1 \over 2} (m_2^2 + {1 \over 2} 
\lambda_3 v^2) D^2 + {1 \over 2} \lambda_1 v h^3 + {1 \over 8} \lambda_1 h^4 
+ {1 \over 8} \lambda_2 D^4 + {1 \over 2} \lambda_3 v h D^2 + {1 \over 4} 
\lambda_3 h^2 D^2,
\end{equation}
where $h$ is the SM Higgs boson. The phenomenology of this simplest of all 
dark-matter scenarios has recently been updated \cite{hllt07,blmrs07}.  Of 
particular interest is the decay $h \to DD$ which is invisible, thus 
allowing \cite{cmr07} $m_h$ to be below the present bound of 114.4 GeV 
from LEP data \cite{lep03}.

In most studies however, dark matter is synonymous with the lightest 
neutralino in $R$-parity conserving supersymmetry \cite{bhs05} which 
again is based on a $Z_2$ discrete symmetry.  In fact, the choice of $Z_2$ 
for dark matter is universally practiced, but is not required by any 
fundamental principle; it is just the simplest hypothesis which works.  
Consider thus the new idea of $Z_3$ dark matter.  How does it work? and 
what are its implications?

Let $\chi$ be a neutral complex scalar singlet, then instead of Eq.~(3), 
consider
\begin{equation}
V = m_1^2 \Phi^\dagger \Phi + m_2^2 \chi^\dagger \chi + {1 \over 2} \lambda_1 
(\Phi^\dagger \Phi)^2 + {1 \over 2} \lambda_2 (\chi^\dagger \chi)^2 + 
\lambda_3 (\chi^\dagger \chi)(\Phi^\dagger \Phi) + {1 \over 6} \mu 
\chi^3 + {1 \over 6} \mu^* (\chi^\dagger)^3.
\end{equation}
This extended Higgs potential is invariant under $Z_3$ with $\chi$ 
transforming as $\omega = \exp(2\pi i/3) = -1/2 + i \sqrt{3}/2$, and 
$\mu$ may be chosen real by absorbing its phase into $\chi$.  With the 
replacement $\Phi^\dagger \Phi = (v+h)^2/2$, this becomes
\begin{eqnarray}
V &=& {1 \over 2} \lambda_1 v^2 h^2 + (m_2^2 + {1 \over 2} \lambda_3 v^2) 
\chi^\dagger \chi + {1 \over 2} \lambda_1 v h^3 + {1 \over 8} \lambda_1 h^4 
+ {1 \over 2} \lambda_2 (\chi^\dagger \chi)^2 \nonumber \\ 
&+& \lambda_3 v h (\chi^\dagger \chi) + {1 \over 2} \lambda_3 h^2 
(\chi^\dagger \chi) + {1 \over 6} \mu [\chi^3 + (\chi^\dagger)^3].
\end{eqnarray}
Comparing this with Eq.~(4), it is clear that $\chi$ is as good a dark-matter 
candidate as $D$, and indistinguishable from $D$ unless the cubic 
self-interaction $\chi^3$ can be established.  This illustrates the 
generic possibility that dark matter may be distinguished from ordinary 
matter by a symmetry larger than $Z_2$ and yet not be discovered easily 
in experiments.

Recalling that dark matter may be related to radiative neutrino mass 
\cite{m06-1}, it is shown below exactly how $\chi$ enables just such 
a scenario in two loops.  There have been two basic two-loop mechanisms of 
radiative neutrino mass \cite{bm89-1}: one via the exchange of two $W$ 
bosons \cite{pt84,bm88,bmp89,bm89-2,cggm94}, the other via a new trilinear 
scalar interaction \cite{z86,b88,ky00,ky01,bm03,mm03,aiky04,ah06}.  Both 
have SM charged leptons in the loop and the new particles involved cannot 
be dark matter.  In a three-loop extension \cite{knt03,cs04}, the innermost 
loop may be populated with dark-matter candidates, but the constraint due to 
flavor changing radiative decays such as $\mu \to e \gamma$ may not be 
so easily satisfied \cite{kms06}.  In the context of the supersymmetric 
$E_6/U(1)_N$ model \cite{m96}, a two-loop mechanism with $Z_2$ dark matter 
has also been proposed recently \cite{ms07}, with the novel feature that the 
$(\lambda_5/2) (\Phi^\dagger \eta)^2$ interaction is generated in one loop.

In the case of $Z_3$ dark matter, consider the following additions to the SM:
\begin{eqnarray}
{\rm scalars:}&& \chi_{1,2,3} ~\sim~ \omega, \\ 
{\rm fermions:}&& (N,E)_{L,R},~S_{L,R} ~\sim~ \omega,
\end{eqnarray}
where $\chi_{1,2,3}$, $N$ and $S$ are neutral and $E$ has charge $-1$.
Using the allowed interactions
\begin{eqnarray}
{\cal L}_{int} &=& h_{ij} (\bar{N}_R \nu_{iL} + \bar{E}_R l_{iL}) \chi_j 
+ f_1 \bar{S}_L (N_R \phi^0 - E_R \phi^+) + f_2 \bar{S}_R (N_L \phi^0 - 
E_L \phi^+) \nonumber \\  &-&  {1 \over 2} f_{3i} \chi_i S_L S_L - {1 \over 2} 
f_{4i} \chi_i S_R S_R - {1 \over 6} \mu_{ijk} \chi_i \chi_j \chi_k + H.c.
\end{eqnarray}
and the allowed Dirac masses $m_N, m_E, m_S$, the following two-loop 
diagram (Fig.~2) for neutrino mass is generated.  Lepton number is now 
conserved multiplicatively with all fields even except $\nu$, $l$, $N$, 
$E$, and $S$ which are odd. Note that $Z_3$ is tailor-made for such a 
mechanism because of the trilinear scalar interaction $\chi^3$.  Note 
also that it is an explicit two-loop realization of the unique 
dimension-five operator $\nu_i \nu_j \phi^0 \phi^0$ \cite{w79,m98}.

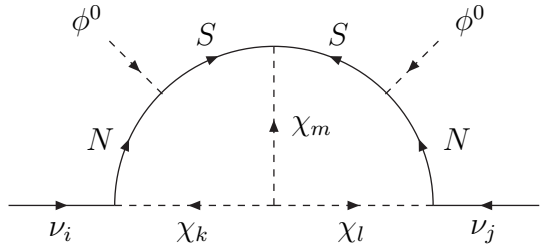
\begin{figure}[htb]
\begin{center}
\begin{picture}(360,120)(0,0)
\ArrowLine(80,10)(120,10)
\DashArrowLine(180,10)(120,10)3
\DashArrowLine(180,10)(240,10)3
\ArrowLine(280,10)(240,10)
\DashArrowLine(180,10)(180,70)3
\DashArrowLine(118,72)(138,52)3
\DashArrowLine(242,72)(222,52)3
\ArrowArcn(180,10)(60,180,135)
\ArrowArcn(180,10)(60,135,90)
\ArrowArc(180,10)(60,45,90)
\ArrowArc(180,10)(60,0,45)

\Text(100,0)[]{$\nu_i$}
\Text(260,0)[]{$\nu_j$}
\Text(150,0)[]{$\chi_k$}
\Text(210,0)[]{$\chi_l$}
\Text(115,35)[]{$N$}
\Text(250,35)[]{$N$}
\Text(195,40)[]{$\chi_m$}
\Text(155,75)[]{$S$}
\Text(205,75)[]{$S$}
\Text(110,80)[]{$\phi^{0}$}
\Text(255,80)[]{$\phi^{0}$}

\end{picture}
\end{center}
\caption{Two-loop generation of neutrino mass.}
\end{figure}

The neutrino mass matrix is then approximately given by
\begin{equation}
({\cal M}_\nu)_{ij} = {v^2 \over 512 \pi^4} \sum_{k,l,m} h_{ik} h_{jl} 
\mu_{klm} \left[ {f_1^2 f_{3m} \over (M_{eff})^2} + {f_2^2 f_{4m} m_N^2 
\over (M'_{eff})^4} \right].
\end{equation}
For illustration, let $h = 0.003$, $f = 0.36$, $\mu = 100$ GeV and 
$M = M' = m_N = 1$ TeV, then neutrino masses are of order 0.1 eV, and flavor 
changing radiative decays such as $\mu \to e \gamma$ (which depend crucially 
on $h$) are well below experimental bounds \cite{bm03}.  [On the other hand, 
there is enough freedom in the choice of the above couplings to allow them 
to be observable in the near future as well.]

Let $\chi_1$ be the lightest of $\chi_{1,2,3}$, then it is a suitable 
dark-matter candidate in the same way as the $Z_2$ candidate $D$ 
\cite{hllt07}.  In addition, $\chi_i$ may be discovered through the decay 
\begin{equation}
E \to l_i \chi_j
\end{equation}
and $\chi_2 \to \chi_1 l^+_i l^-_j$, etc.  Another important feature of 
this model is the mixing between $S$ and $N$ through $\langle \phi^0 
\rangle = v/\sqrt{2}$.  This allows the decay of the heavier mass eigenstate 
$N_2$ into the lighter $N_1$, i.e.
\begin{equation}
N_2 \to N_1 Z,
\end{equation}
and $N_1 \to \nu_i \chi_j$.  The decay $E \to N_1 W^-$ may also be possible.
In searching for the SM Higgs boson $h$, the invisible decay $h \to \chi_1 
\chi_1$ and the more generic $h \to \chi_i \chi_j$ should also be 
kept in mind.

A possible variant of this model is to substitute the particles of Eqs.~(7) 
and (8) with
\begin{eqnarray}
{\rm scalars:}&& \chi, ~(\eta^+,\eta^0) ~\sim~ \omega, \\ 
{\rm fermions:}&& (N_{1,2,3})_{L,R} ~\sim~ \omega,
\end{eqnarray}
where $N$ is neutral.  In this case, the two-loop diagram for neutrino mass 
(Fig.~3) involves necessarily the mixing of $\chi$ and $\eta^0$.  This 
renders them unsuitable as dark-matter candidates because each eigenstate 
must couple to the $Z$ boson (in the same way as the scalar neutrino in 
supersymmetry) and thus ruled out by direct-search experiments \cite{bhs05}. 
In the case of $Z_2$ dark matter, $\eta^0$ is allowed to be split so that 
$\eta_R^0$ and $\eta^0_I$ have different masses, thereby evading the above 
constraint.  Here this is not possible because of the $Z_3$ symmetry.

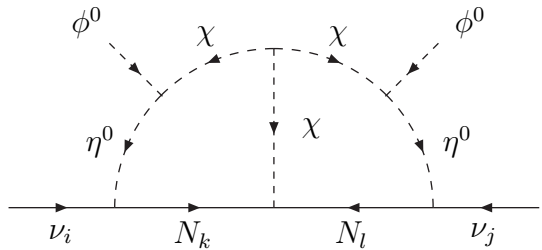
\begin{figure}[htb]
\begin{center}
\begin{picture}(360,120)(0,0)
\ArrowLine(80,10)(120,10)
\ArrowLine(120,10)(180,10)
\ArrowLine(240,10)(180,10)
\ArrowLine(280,10)(240,10)
\DashArrowLine(180,70)(180,10)3
\DashArrowLine(118,72)(138,52)3
\DashArrowLine(242,72)(222,52)3
\DashArrowArc(180,10)(60,135,180)3
\DashArrowArc(180,10)(60,90,135)3
\DashArrowArcn(180,10)(60,90,45)3
\DashArrowArcn(180,10)(60,45,0)3

\Text(100,0)[]{$\nu_i$}
\Text(260,0)[]{$\nu_j$}
\Text(150,0)[]{$N_k$}
\Text(210,0)[]{$N_l$}
\Text(115,35)[]{$\eta^0$}
\Text(250,35)[]{$\eta^0$}
\Text(195,40)[]{$\chi$}
\Text(155,75)[]{$\chi$}
\Text(205,75)[]{$\chi$}
\Text(110,80)[]{$\phi^{0}$}
\Text(255,80)[]{$\phi^{0}$}

\end{picture}
\end{center}
\caption{Another two-loop generation of neutrino mass.}
\end{figure}

The lightest $N_i$ may be considered \cite{m06-1,kms06,ks06} as the 
dark-matter candidate in this case, but because its relic abundance depends 
on its interaction with charged leptons, flavor-changing radiative decays 
such as $\mu \to e \gamma$ are difficult to suppress.  One solution 
\cite{bm07} is to add a singlet scalar which is also trivial under $Z_3$ 
so that $N \bar{N} \to hh$ occurs at tree level.

In $R$-parity conserving supersymmetry, the dark-matter candidate, i.e. 
the lightest neutralino, is just one of an entire class of new particles 
to be discovered which are odd under $Z_2$.  In the context of $Z_3$ dark 
matter, this may also be the case.  If they have $SU(3)_C$ interactions, 
they will also be produced abundantly at the LHC and have a good chance 
of being discovered in the near future.

Another possible avenue of thought is that dark matter may be a hint that 
there could be new particles in a hidden sector which communicate with 
ordinary particles only through the Higgs sector.  This makes sense 
also in the context of the Minimal Supersymmetric Standard Model, 
where the only allowed bilinear term of the superpotential is $\mu H_u H_d$. 
If $\mu$ is replaced by a singlet superfield, the latter may serve as 
the link to a completely new world of particles, perhaps with its own gauge 
group and conservation laws.  There may even be several such links.
 
In conclusion, it has been pointed out that the universal assumption of 
$Z_2$ dark matter is not the only available option.  A specific $Z_3$ 
alternative is proposed which coincides with the new notion that dark matter 
is responsible for neutrino mass, generating the latter radiatively in two 
loops. The dark-matter candidate here is a complex singlet scalar field which 
interacts with ordinary matter mostly through the canonical Higgs boson of 
the Standard Model.

This work was supported in part by the U.~S.~Department of Energy under Grant 
No.~DE-FG03-94ER40837.

\newpage

\baselineskip 18pt

\bibliographystyle{unsrt}

\end{document}